\documentclass[reprint,english,aps,prb,twocolumn,amsmath, amssymb, showpacs]{revtex4}
\usepackage[T1]{fontenc}
\usepackage{subfigure}
\usepackage[latin1]{inputenc}
\usepackage{graphicx}
\usepackage{epstopdf}
\usepackage{epsfig}
\usepackage{prettyref}
\usepackage{babel}
\usepackage{color}
\makeatother

\begin{document}

\title{A critical examination of magnetic states of La$_{0.5}$Ba$_{0.5}$CoO$_3$: non-Griffiths phase and interacting ferromagnetic-clusters }

\author{Devendra Kumar}
\email{deven@csr.res.in}
\affiliation{UGC-DAE Consortium for Scientific Research, University Campus, Khandwa Road, Indore-452001,India}

\author{A. Banerjee}
\affiliation{UGC-DAE Consortium for Scientific Research, University Campus, Khandwa Road, Indore-452001,India}

\begin{abstract}
We report detailed dc magnetization, linear and non-linear ac susceptibility measurements on the hole doped disordered cobaltite La$_{0.5}$Ba$_{0.5}$CoO$_3$. Our results show that the magnetically ordered state of the system consists of coexisting non-ferromagnetic phases along with percolating ferromagnetic-clusters. The percolating ferromagnetic-clusters possibly undergo a 3D Hisenberg like magnetic ordering at the Curie temperature of 202(3)~K. In between 202 and 220~K, the linear and non-linear ac susceptibility measurements show the presence of magnetic correlations even when the spontaneous magnetization is zero which indicates the presence of preformed short range magnetic-clusters. The characteristics of these short range magnetic-clusters that exist above Curie temperature are quite distinct than that of Griffiths phase e.g the inverse dc susceptibility exhibits an field independent upward deviation, and the second harmonic of ac susceptibility is non-negative. Below Curie temperature the system exhibit spin-glass like features such as irreversibility in the field cooled and zero field cooled magnetization and frequency dependence in the peak of ac susceptibility. The presence of a spin or cluster -glass like state is ruled out by the absence of field divergence in third harmonic of ac susceptibility and zero field cooled memory. This indicates that the observed spin-glass like features are possibility due to progressive thermal blocking of ferromagnetic-clusters which is further confirmed by the
Wohlfarth's model of superparamagnetism. The frequency dependence of the peak of ac susceptibility obeys the Vogel-Fulcher law with $\tau_0\approx10^{-9}$s. This together with the existence of an AT line in H-T space indicates the existence of significant inter-cluster interaction among these ferromagnetic-clusters.
\end{abstract}



\maketitle

\section{Introduction}
The transition metal oxides e.g. manganites, cuprates, and cobaltites exhibit complex phase diagram including the microscopically inhomogeneous electronic states due to interplay of various competitive electronic energies such as electron kinetic energy, electron-electron coulomb repulsion, spin-spin, spin-orbit, and crystal field interactions.\cite{Dagotto, Rini} Of these oxides, the cobaltite LaCoO$_3$ exhibits a unique property of temperature and doping dependent spin state transition.\cite{Louca1, Podlesnyak} The Co$^{3+}$ ion in LaCoO$_3$ can exist in low spin (LS) state with configuration t$_{2g}^6$e$_g^0$ ($S$=0), intermediate spin (IS) state with configuration t$_{2g}^5$e$_g^1$ ($S$=1), and high spin (HS) state with configuration t$_{2g}^4$e$_g^2$ ($S$=2). The LaCoO$_3$ have a charge transfer insulator type non-magnetic  ground state with Co$^{3+}$ ion in the LS state; it starts showing magnetic moment above 30~K and exhibits a paramagnetic like behavior above 100~K.\cite{Raccah, Kriener} This change in magnetic moment and behavior is attributed to thermally driven spin state transition of Co$^{3+}$ ion,  but the nature of transition whether it is a LS-IS transition or LS-HS transition is still not completely settled.\cite{Raccah, Kriener, Zobel, Noguchi, Korotin, Haverkort, Radaelli, Podlesnyak} The hole doping of LaCoO$_3$ by replacing the trivalent La$^{3+}$ with divalent Sr$^{2+}$ or Ba$^{2+}$ generates Co$^{4+}$, and each of these Co$^{4+}$ transforms their six nearest Co$^{3+}$ neighbors into the IS state by forming octahedrally shaped spin-state polarons.\cite{Louca, Phelan, Podlesnyak1} In these polarons, e$_g$ electrons of Co$^{3+}$ are delocalized and are shared by Co$^{3+}$ and Co$^{4+}$ ions of the polaron, while t$_{2g}$ electrons of both the ions are localized and couple ferromagnetically via double exchange interaction. For small hole doping
these isolated spin state polarons are stable within the nonferromagnetic matrix. Additional hole doping enhances the number density of spin state polarons, and above a critical doping of x=0.04, the enhanced polaron density  causes a decay of polaronic state due to ferromagnetic (FM) interaction between the intra-polaronic Co$^{3+}$ ions at the cost of the antiferromagnetic (AFM) intra-polaronic interaction.\cite{Podlesnyak2} This, in turn, results in the formation of hole rich ferromagnetic spin clsuters embedded in non-ferromagnetic insulating matrix. On further enhancing the hole doping, at a critical concentration ($x$=0.18 for Sr$^{2+}$ and $x$=0.2 for Ba$^{2+}$), the ferromagnetic metallic clusters eventually percolates giving rise to long range ferromagnetic ordering and metallic conductivity.\cite{Kriener1}

For La$_{1-x}$Sr$_{x}$CoO$_3$, a spin-glass like state is observed below the critical doping concentration for percolation of ferromagnetic metallic regions and above this a ferromagnetic or a ferromagnetic-cluster state is reported.\cite{Kriener1, Mandal, Wu, Mukherjee, Samal}
But a recent report on La$_{0.5}$Sr$_{0.5}$CoO$_3$ show the presence of glassy dynamics even in the so called ferromagnetic or ferromagnetic-cluster state which has been attributed to coexistence of spin or cluster -glass like phase along with  percolating ferromagnetic-clusters.\cite{Samal1} The absence of exchange bias effect in La$_{0.5}$Sr$_{0.5}$CoO$_3$ clearly indicates that the spin or cluster -glass like phase is not present at the interface of ferromagnetic-clusters and non-ferromagnetic matrix, but instead, it probably coexist as small patches along with the percolating backbone of ferromagnetic-clusters. The nature of magnetic state in La$_{1-x}$Ba$_{x}$CoO$_3$ with Ba$^{2+}$ having a larger ionic radii than Sr$^{2+}$ (ionic radii of La$^{3+}$=1.216\AA, Ba$^{2+}$=1.47\AA, Sr$^{2+}$=1.31\AA) is relativity less studied, and early reports indicate the presence of ferromagnetic-metallic ground state for $x>0.2$.\cite{Kriener1, Mandal} The higher ionic radii of Ba$^{2+}$ (a) enhances the local randomness due to larger size mismatch between the Ba$^{2+}$ and La$^{3+}$ ions, and (b) reduces the overall distortion from ideal pervoskite structure and so the tolerance factor $t$ approaches to 1. This enhancement in tolerance factor straightens the Co-O-Co bonds which in turn increases the ferromagnetic coupling due to double exchange interaction between Co$^{3+}$  and Co$^{4+}$  ions. Furthermore, the Ba$^{2+}$ doping enhances the concentration of Jahn-Teller (J-T) active IS state because of lattice expansion and the formation of J-T magnetopolaron is found to be most preferable for Ba doped cobaltites.\cite{Phelan, Phelan1}

In this paper we present a detailed study of La$_{0.5}$Ba$_{0.5}$CoO$_3$ with an aim to understand its different magnetic states. We have compared our results with that of La$_{0.5}$Sr$_{0.5}$CoO$_3$ to bring out possible effect of differences in ionic radii of Ba$^{2+}$ and La$^{2+}$.
Our results show that the magnetically ordered state of La$_{0.5}$Ba$_{0.5}$CoO$_3$ consist of non-ferromagnetic phases coexisting along with percolating backbone of ferromagnetic-clusters. These ferromagnetic-clusters have critical exponent $\gamma$=1.29(1) indicating the possibility of 3D Hisenberg like spin ordering at the Curie temperature ($T_C$) of 202(3)~K. Above $T_C$, short range magnetic-clusters with characteristics quite different from Griffiths phase exists up to around 220~K. Below $T_C$ the magnetic state of the system exhibits spin-glass like behavior, but in contrast to La$_{0.5}$Sr$_{0.5}$CoO$_3$, this behavior does not originate from coexistence of spin or cluster -glass like phases along with percolating ferromagnetic-clusters. Furthermore, our analysis show that the spin-glass like dynamics in La$_{0.5}$Ba$_{0.5}$CoO$_3$ is due to superparamagnetic like thermal blocking of the dynamics of interacting ferromagnetic-clusters.

\section{Experimental Details}
Polycrystalline La$_{0.5}$Ba$_{0.5}$CoO$_3$ samples are prepared by pyrophoric method~\cite{Pati} using high purity (99.99\%) La$_2$O$_3$, BaCoO$_3$, and Co(NO$_3$)$_2$6H$_2$O. The stichometric ratio of La$_2$O$_3$, BaCoO$_3$, and Co(NO$_3$)$_2$6H$_2$O are separately dissolved in dilute nitric acid and then these solutionis are mixed with the triethanolamine (TEA) keeping the pH highly acidic. The final solution is dried at ~100~$^\circ$C, which burns and yields a black powder that is palletized and subsequently annealed at 1100~$^\circ$C for 12~hour. These samples are characterized by X-Ray diffraction on a Bruker D8 Advance X-ray diffractometer using Cu-K$\alpha$ radiation. The dc magnetization measurements are performed on a 14~T Quantum Design physical property measurement system-vibrating sample magnetometer and the low field ac susceptibility measurements are carried out on a ac-susceptibility setup which is described in Reference~\onlinecite{Bajpai}.

\begin{figure}[!t]
\begin{centering}
\includegraphics[width=0.8\columnwidth]{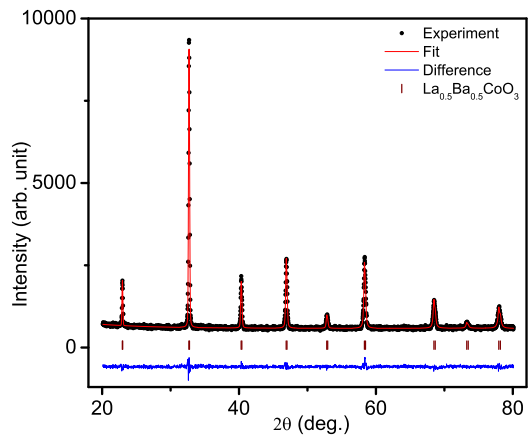}
\par\end{centering}
\caption{Room temperature X-ray diffraction pattern of La$_{0.5}$Ba$_{0.5}$CoO$_3$. The solid circles show the experimental X-ray diffraction data, the red line on the experimental data show the Rietveld refinement for simple cubic Pm-3m structure with $\chi^2$=1.34, the short vertical lines give the Bragg peak positions, and the bottom blue line gives the difference between the experimental and calculated pattern.} \label{fig: XRD}
\end{figure}

The X-ray diffraction data of La$_{0.5}$Ba$_{0.5}$CoO$_3$ is collected at room temperature and analyzed with Rietveld structural refinement using FULLPROF software.\cite{Carvajal} Figure \ref{fig: XRD} show the XRD data, the Rietveld fit profile, the Bragg positions, and the difference in experimental and model results. The Rietveld refinement show that the sample is single phase and crystallizes in simple cubic Pm-3m structure with lattice constant $a$=3.8726(2)\AA{} and  unit cell volume $V$=58.078(4)\AA$^3$. The unit cell volume in disordered cobaltites depends on the oxygen stoichiometry, and the comparison of  our result with that of Reference~\onlinecite{Troyanchuk} suggests that the oxygen non-stoichiometry is much less than 0.05. The average crystallite size of the sample is estimated from XRD data using the Scherrer formula which comes around 85~nm.

\section{Results and Discussions}
\subsection{DC Magnetization}
\subsubsection{Thermomagnetic irreversibility}
\begin{figure}[!t]
\begin{centering}
\includegraphics[width=0.8\columnwidth]{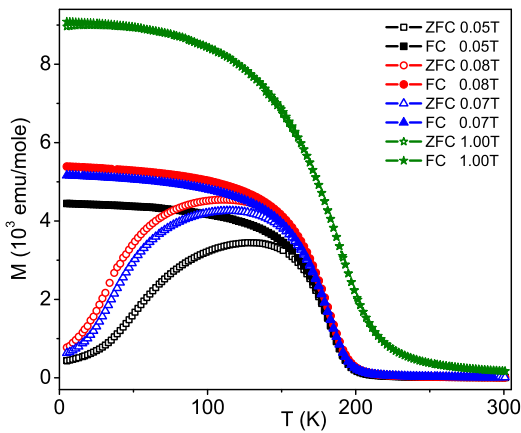}
\par\end{centering}
\caption{Temperature dependence of  magnetization under FC and ZFC protocol at various measuring fields.} \label{fig: MvsT}
\end{figure}

Figure \ref{fig: MvsT} show the temperature variation of magnetization in field cooled (FC) and zero field cooled (ZFC) protocol. In the FC protocol, the sample is cooled to 5~K in presence of measuring field and the magnetization is recorded in heating run keeping the field constant. In ZFC protocol the sample is cooled to 5~K in zero field and then the measuring field is applied and magnetization is recorded as a function of temperature in the heating run. On lowering the temperature, around 200~K,  both the FC and ZFC magnetization curves show a rapid increase in magnetization which is indicative of paramagnetic to ferromagnetic transition. On further lowering the temperature, the FC curve keeps evolving while the ZFC curve bifurcates with that of FC at the temperature T$_{irr}$ and exhibits a broad peak at a temperature $T_p$. On increasing the measuring field $T_{irr}$ and $T_p$ decreases with an enhancement in broadening of ZFC peak. At 1~T, the FC and ZFC curves almost coincide. The bifurcation in FC-ZFC magnetization along with a peak in ZFC magnetization indicates about the presence of a spin-glass,\cite{Maydosh} cluster-glass,\cite{Deac, Huang} super-paramagnetic,\cite{Knobel, Pramanik} or ferromagnetic state.\cite{Anil} At low fields, $T_p$ < $T_{irr}$, and below $T_p$ the FC magnetization is not constant with temperature. This observation is not in agreement with that of canonical spin-glasses and suggests that the system is possibly in a  cluster-glass, super-paramagnetic, or ferromagnetic state. Similar observations has been made on the other relatively well studied half doped disordered cobaltite La$_{0.5}$Sr$_{0.5}$CoO$_3$.\cite{Wu, Samal, Samal1}

\subsubsection{Non-Griffiths phase}
\begin{figure} [!t]
\begin{centering}
\includegraphics[width=0.8\columnwidth]{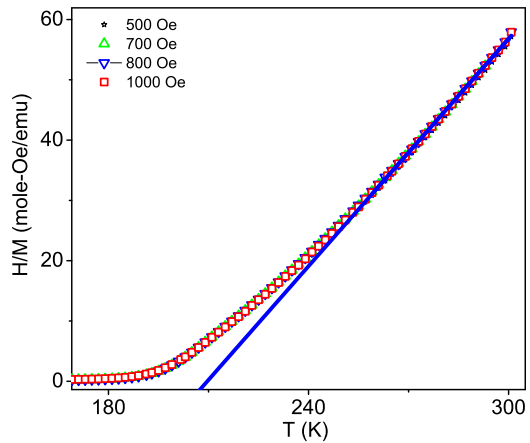}
\par\end{centering}
\caption{H/M versus temperature at 500, 700, 800, and 1000~Oe. The solid line show the fitting of Curie-Weiss equation.} \label{fig:CW}
\end{figure}

Figure \ref{fig:CW} displays the temperature variation of inverse dc susceptibility (H/M) at 500, 700, 800, and 1000~Oe taken under the FC protocol. The H/M fits well with the Curie-Weiss law at high temperatures (T>250~K) with Curie constant $C$=1.59(1)~emu-K/mole-Oe and Weiss constant $\theta$=210(2)~K. The Curie constant gives an effective value of paramagnetic moment $\mu_{eff}=$ 3.566(3)~$\mu_B$/f.u. and the positive value of Weiss constant indicates the dominance of ferromagnetic correlations in the ordered state. The inverse dc susceptibility exhibits a upward deviation from the Curie-Weiss law at a temperature $T^*$($\approx$ 250~K) which is greater than $T_C$, and this deviation is found to be field independent. 
This upward deviation of the inverse dc susceptibility from Curie-Weiss law in La$_{0.5}$Ba$_{0.5}$CoO$_3$ is quite distinct from that of Griffiths phase seen in many randomly doped transition metal oxides, where the presence of short range ferromagnetic correlations above $T_C$ gives a field dependent downward deviation in the inverse dc susceptibility.\cite{Griffiths, Magen, Pramanik1} A similar non-Griffiths like behavior in the inverse dc susceptibility is also observed in La$_{1-x}$Sr$_x$CoO$_3$.\cite{Caciuffo, He} In case of La$_{1-x}$Sr$_x$CoO$_3$, the small angle neutron diffraction measurements exhibit a sharp onset in the spin correlation at the temperature $T^*$ (where the inverse susceptibility deviates from Curie Weiss law) which has been interpreted as the emergence of short range ferromagnetic clusters.\cite{He} Above $T_C$, the  deviation from Curie Weiss law in inverse dc susceptibility can be (a) because of composition fluctuation and so having regions with Curie temperature higher than the bulk of the sample or (b) because of the presence of short range ferromagnetic correlations with zero spontaneous magnetization as is the case for Griffiths phase. The possible mechanism for upward deviation in inverse dc susceptibility, whether it is due to composition fluctuation or due to existence of short range ferromagnetic correlations, will be probed by the Arrot plot and ac susceptibility measurements in section \ref{subsubsection:isothermal magnetization} and \ref{subsubsection:non-Griffiths} respectively.

\subsubsection{Coexistence of ferromagnetic and non-ferromagnetic phases}
\label{subsubsection:isothermal magnetization}
\begin{figure} [!t]
\begin{centering}
\includegraphics[width=0.8\columnwidth]{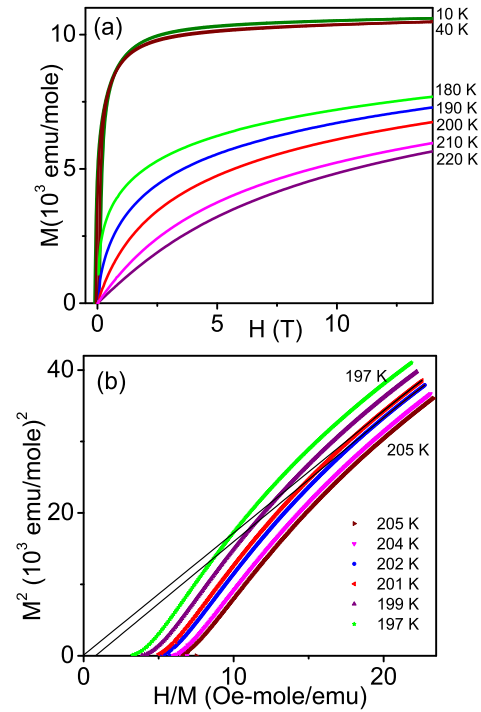}
\par\end{centering}
\caption{(a) Magnetization versus field at 10~K, 40~K, 180~K, 190~K, 200~K, 210~K, and 220~K (b) Arrot plot (M$^2$ vs. H/M) of the magnetization isotherms at 197~K, 199~K, 201~K, 202~K, 204~K, and 205~K. The solid black lines are straight line fit to M$^2$ vs. H/M curve at high field which are extrapolated to H=0.} \label{fig:MH}
\end{figure}

In figure \ref{fig:MH} (a) we show the magnetization versus field plot at 10~K, 40~K, 180~K, 190~K, 200~K, 210~K, and 220~K. At low temperatures, for example at 10~K the magnetization exhibits a saturation like behavior at high magnetic fields which is typical of a ferromagnet, but a careful observation of the data indicates the presence of a non-saturating magnetization along with the saturating ferromagnetic component. The presence of this non-saturating component prohibits the magnetization from saturating even at high magnetic fields. The existence of a non-ferromagnetic component along with the ferromagnetic component is in agreement with the cluster model of other disordered cobaltite La$_{1-x}$Sr$_x$CoO$_3$ where a number of studies have shown the presence of non-ferromagnetic Co$^{3+}$ matrix with antiferromagnetic interactions that coexist along with ferromagnetic-clusters.\cite{Wu, Samal} The ferromagnetic component can be extracted from the total magnetization by assuming that the total magnetization can be written as $M_{tot}$=$M_F$ + $\chi_{AF}$H where $M_F$ is the saturation value of ferromagnetic component and $\chi_{AF}$ is the slope of M vs. H curve at high fields. Using this to fit the magnetization versus field curve above 12~T at 10~K, we estimate the saturation magnetization of ferromagnetic component as 1.855(1)~$\mu_B$/Co. The value of saturation magnetization of ferromagnetic-clusters is smaller than that expected from the spin only value ($M_s=gS\mu_B=2.5\mu_B$) when both the Co$^{3+}$ ($S$=1) and Co$^{4+}$ ($S$=3/2) are in IS state. It is to be noted that the similar results about the difference in experimental and expected saturation magnetization has also been reported on La$_{1-x}$Sr$_x$CoO$_3$.\cite{Wu, Samal} On the basis of the band structure calculations, Ravindra $et$~$al.$\cite{Ravindran} have shown that the hole doping in these materials reduces the ionicity, enhances the Co-O hybridization, and stabilizes the IS state. Due to enhanced Co-O hybridization the expected average Co moment is reduced compared to the prediction of simple ionic model.

In figure \ref{fig:MH} (b), the M-H isotherms around 200~K are plotted as M$^2$ versus H/M which is known as Arrot plot.\cite{Arrott} In these plots, the intercept of the linear fitting of high field data on the X and Y axis gives inverse susceptibility and spontaneous magnetization respectively and the one passing through origin gives the ferromagnetic transition temperature T$_C$. At 201~K, the value of spontaneous magnetization is 0.058~$\mu_B$/Co which shows the presence of ferromagnetic interactions, and  the line passing through origin will correspond to M$^2$ versus H/M curve lying in between 201-202~K indicating that the T$_C$ lies in between. Above T$_C$, e.g. at 204~K and 205~K, the spontaneous magnetization (M$_S$) is zero indicating the absence of long range ferromagnetic ordering.

\subsection{AC Susceptibility}
In order to get a better understanding of the magnetically ordered state, we have performed ac susceptibility measurements at low fields which probe the dynamics of the system at the time scales decided by the measuring frequency range. The magnetization ($M$) of a system can be expressed in terms of the applied field ($H$) as:
\begin{equation}
M(H)=M_0+\chi_1H+\chi_2H^2+\chi_3H^3+...\label{eq:nonlin}
\end{equation}
where $M_0$ is the spontaneous magnetization, $\chi_1$ is the linear susceptibility and $\chi_2$, $\chi_3$,.. are the nonlinear susceptibilities which can be identified with the Taylor series expansion of $M(H)=M_0+(1/1!)(dM/dH)_{H=0}H+(1/2!)(d^2M/dH^2)_{H=0}H^2+..$ .
\subsubsection{Nature of magnetically ordered state}
\begin{figure} [!t]
\begin{centering}
\includegraphics[width=0.8\columnwidth]{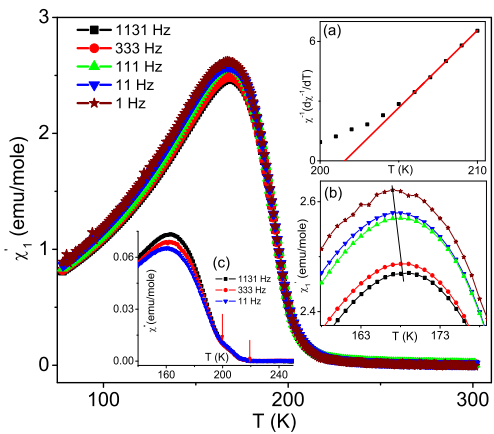}
\par\end{centering}
\caption{Temperature dependence of the real part of linear ac susceptibility at various frequencies at the field of 2.21~Oe. The inset (a) show the $\chi^{-1}d\chi^{-1}/dT$ versus T at 131~Hz at ac field of 0.57~Oe, the inset (b) show the expanded view of the peak in ac susceptibility and the inset (c) show the imaginary part of linear ac susceptibility.} \label{fig: ACR}
\end{figure}

Figure \ref{fig: ACR} show the real part of linear ac susceptibility ($\chi^{'}_1$) measured in the ac field of 2.21~Oe and frequency 1131, 333, 131, 11, and 1~Hz. The $\chi^{'}_1$ exhibits a broad peak similar to that of ZFC magnetization and the peak position ($T_B$) in $\chi^{'}_1$ decreases on increasing the measuring frequency (see inset (b) of figure \ref{fig: ACR}) which is a common feature of spin-glass, cluster-glass, and superparamagnetic systems; and the presence of frequency dependence in $T_B$ clearly rules out the possibility of normal long range ferromagnetic state. The frequency dependence in $\chi^{'}_1$ is quantified as $\Phi=\Delta T_B/(T_B\Delta$log$_{10}f)$ and the estimated value of $\Phi$ is 0.0023 which is in agreement with typical values seen in canonical spin-glasses, cluster-glasses, superspin-glass, or interacting super-paramagnets (0.02-0.005),\cite{Goya, Cong, Thakur} and two order of magnitude lower than that observed in noninteracting super-paramagnets (0.1-0.3).\cite{Toro, Dormann}

\begin{figure} [!t]
\begin{centering}
\includegraphics[width=0.8\columnwidth]{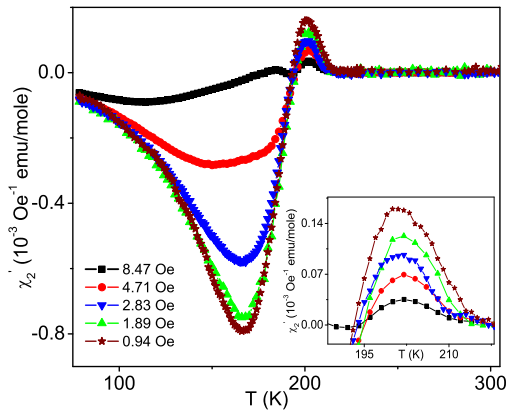}
\par\end{centering}
\caption{Temperature dependence of the real part of second harmonic of AC susceptibility at various fields at 131~Hz. The inset show the expanded view of the small peak at 202~K.} \label{fig: chi2}
\end{figure}

In absence of the time inversion symmetry breaking field, $M(H)$=-$M(-H)$, all the even terms in equation \ref{eq:nonlin} i.e. $\chi_2$, $\chi_4$ are zero. The $\chi_2$ is observed in presence of a superimposed external dc field or an internal field which originates from magnetically correlated spins. For canonical spin-glass the coefficient of even powers of $H$ in equation \ref{eq:nonlin} are zero. The real part of nonlinear susceptibility $\chi_2$ is plotted in figure \ref{fig: chi2}. $\chi_2$ is zero in paramagnetic phase, has a small positive peak at 202~K, then a large negative peak around 167~K, and thereafter it slowly approaches to zero. Both the positive and negative peaks in $\chi_2$ diminishes on increasing the ac field. Below T$_C$, the negative value of $\chi_2$ clearly show the presence of ferromagnetic ordering which also rules out the presence of canonical spin-glass state, but the possibility of coexistence of spin-glass phase along with ferromagnetic-clusters remains open. Moreover, Kouvel-Fisher (K-F) analysis of the ac susceptibility data is performed to get an accurate measure of $T_C$ and an idea about the nature of ferromagnetic transition. According to K-F, the zero field susceptibility in the vicinity of $T_C$ varies as\cite{Kouvel}
\begin{equation}
Y=\left[\frac{d}{dT}(\textrm{ln}\chi_0^{-1})\right]^{-1}=\frac{(T-T_C)}{\gamma}\label{eq:nonlin}
\end{equation}
In the inset~(a) of figure \ref{fig: ACR} we have plotted the $Y$ versus temperature for 131~Hz at the AC field of 0.57~Oe in the vicinity of $T_C$. The temperature range of straight line fitting is selected in such a way that fitting of any subset of the data gives the same parameters (within the error bar). The inverse of the slope and the intercept on $T$ axis gives $\gamma$=1.29(1) and $T_C$=202(3)~K respectively. The $T_C$ value is in agreement with that obtained from Arrot plot in section \ref{subsubsection:isothermal magnetization}.  The $\gamma$ value lies between the 3D Ising model (1.241) and the 3D Hisenberg model (1.386) and are in close agreement with the values reported for La$_{0.67}$Sr$_{0.33}$CoO$_3$ ($\gamma$=1.310(1)), La$_{0.5}$Sr$_{0.5}$CoO$_3$ ($\gamma$=1.27(2)) and conventional  ferromagnet Ni ($\gamma$=1.34(1)).\cite{Menyuk, Mukherjee, Khan} The La$_{0.67}$Sr$_{0.33}$CoO$_3$ is believed to order as 3D Hisenberg model and therefore a similar ordering is also expected for La$_{0.5}$Ba$_{0.5}$CoO$_3$

\subsubsection{Further evidence of non-Griffiths phase}
\label{subsubsection:non-Griffiths}
From figure \ref{fig: chi2} we note that the $\chi_2$ starts to have a non zero value below 220~K, a temperature which is higher than the $T_C$  obtained from Kouvel-Fisher (K-F) analysis. Also, the imaginary part of the first harmonic of ac susceptibility ($\chi^{''}$)  starts showing a non zero value below $\approx$ 220~K (see inset (c) of figure \ref{fig: ACR}). These two observations clearly show the presence of magnetic correlations much above $T_C$ where the spontaneous magnetization is zero (see figure \ref{fig:MH} (b)). The absence of spontaneous magnetization establish un-ambiguously that the non zero value of $\chi_2$ and $\chi^{''}$ are not because of some ferromagnetic regions with $T_C$  higher than the bulk of the system due to composition fluctuation. This suggests that the presence of magnetic correlation above $T_C$ is due to the preformation of short range magnetic-clusters at a temperature $T^*$ much above $T_C$ and so the observed upward deviation from the Curie Weiss law in inverse dc susceptibility possibly owes its origin to the existence of short range magnetic-clusters.
We note that while in small angle neutron diffraction measurements of La$_{1-x}$Sr$_x$CoO$_3$ the ferromagnetic correlations appears at the $T^*$,\cite{He} the non-zero value of $\chi_2$ and $\chi^{''}_1$ in ac susceptibility of La$_{0.5}$Ba$_{0.5}$CoO$_3$ is detectable only below 220~K which is lower than the $T^*$($\approx$250~K). In the preformed magnetic-cluster state, we get a positive value of $\chi_2$ which changes sign on lowering the temperature below $T_C$ (see figure \ref{fig: chi2}). This is quite different from Griffiths phase where a negative value of $\chi_2$ is observed at all $T$ below the Griffiths temperature.\cite{Pramanik1}

In La$_{1-x}$Sr$_x$CoO$_3$, the upward deviation of inverse dc susceptibility in presence of short range ferromagnetic-clusters ($T_C$<$T$<$T^*$) is argued to be possibly due to antiferromagnetic interactions which may favor antiparallel alignment of these clusters resulting in suppression of susceptibility.\cite{He}
The existence of short range ferromagnetic-clusters above $T_C$ and how these ferromagnetic-clusters give a field independent upward deviation in inverse dc susceptibility is not yet properly understood. Here we show the existence of non-Griffiths like phase in  La$_{0.5}$Ba$_{0.5}$CoO$_3$ which in combination with the results of La$_{1-x}$Sr$_x$CoO$_3$ suggest that the non-Griffiths like phase may also be present in other hole doped disordered cobaltites. Our results gives a deeper insight about the characteristics of such non-Griffiths phase which would be important in developing a definitive understanding of them.

\begin{figure} [!t]
\begin{centering}
\includegraphics[width=0.8\columnwidth]{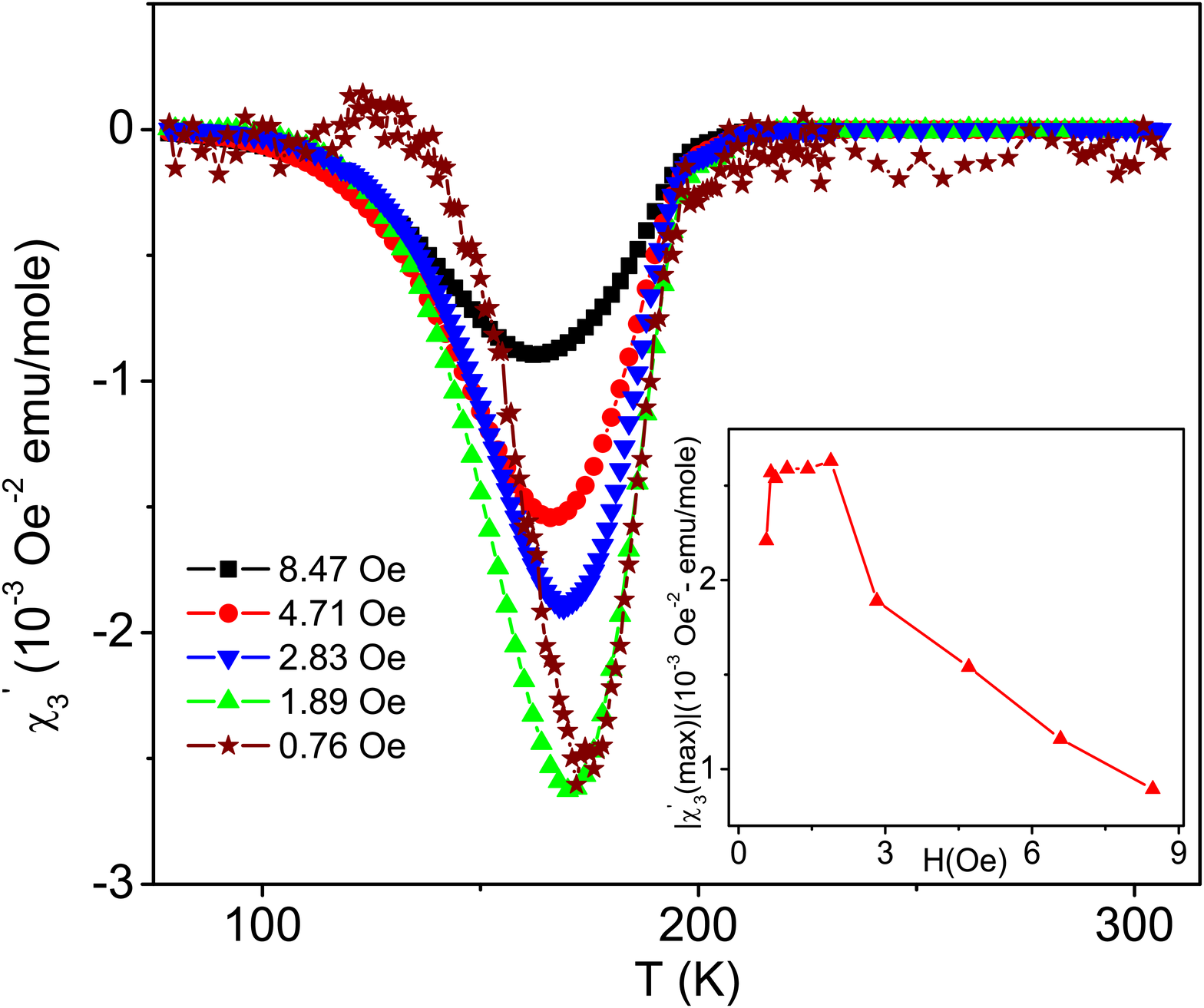}
\par\end{centering}
\caption{Temperature dependence of the real part of third harmonic of ac susceptibility at various fields at 131~Hz. The inset show the ac field dependence of the peak value of $\chi_{3}$ which is $\chi_{3}^{'}(max))$.} \label{fig: chi3}
\end{figure}

\subsubsection{Absence of spin or cluster -glass like transition}
After discarding the possibility of canonical spin-glass state on the basis of non zero value of $\chi_2$
we need to identify the origin of frequency dependence in $T_B$ from remaining possibilities, which are the cluster-glass, the super-paramagnetism, and the coexistence of non-ferromagnetic glassy (e.g. spin or cluster-glass type) phases along with the ferromagnetic clusters. In the first two cases, relaxing entities are the superspins i.e. the moment of a single magnetic domain (cluster) instead of atomic spins and the maximum feasible size of the domain (cluster) is that of a crystallite; while for the last case, as reported for La$_{0.5}$Sr$_{0.5}$CoO$_3$, the relaxing entities are not well understood and the atomic-spins, spin-clusters are among the various possibilities. It is quite difficult to distinguish whether the slowing down in spin dynamics is due to progressive thermal blocking or due to spin-glass like cooperative freezing of the fluctuating entities.
To determine the nature of spin dynamics, we have measured the third harmonic of ac susceptibility ($\chi_3$) which is proportional (and opposite in sign) to spin-glass susceptibility ($\chi_{SG}$). The negative divergence of $\chi_3$ at $T_g$ in the limit of $H\to0$ gives the direct evidence of spin-glass like critical slowing down of the fluctuating entities and hence confirms unambiguously the presence of spin or cluster -glass phase.\cite{Suzuki, Bajpai1} The temperature dependence of the real part of the third harmonic of ac susceptibility at 131~Hz and at different ac fields is plotted in figure~\ref{fig: chi3}. The magnitude of the peak in $\chi_3$ ($\chi_{3}^{'}(max))$) depends on the ac field and the field dependence of $\chi_{3}^{'}(max))$ is plotted in the inset of figure \ref{fig: chi3}. The $\chi_{3}^{'}(max))$ does not diverge as $H\to0$ which clearly shows that fluctuating entities does not freeze in a spin or cluster -glass state. We do not observe any ZFC memory effect which further supports the absence of spin or cluster -glass like freezing in the system.
These results suggest that the observed frequency dependence in $\chi_1$ is possibility due to progressive thermal blocking of fluctuating entities.

\begin{figure} [!t]
\begin{centering}
\includegraphics[width=0.8\columnwidth]{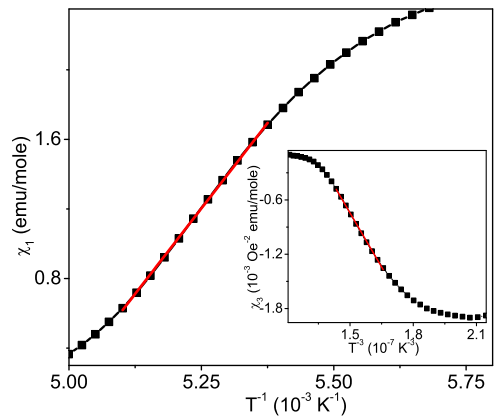}
\par\end{centering}
\caption{$\chi^{'}_1$ versus $T^{-1}$ above the blocking temperature. The inset show the $\chi^{'}_3$ versus $T^{-3}$ above the blocking temperature. The straight lines are the fitting of equation \ref{eq:Wohlfarth1} and \ref{eq:Wohlfarth2} to the data.} \label{fig: superpara}
\end{figure}

\subsubsection{Superparamagnetic behavior of ferromagnetic clusters}
The existence of super-paramagnetism behavior, i.e. progressively thermal blocking of single domain magnetic clusters is further substantiated by Wohlfarth's model of superparamagnets  which shows that the magnetization of an ensemble of magnetic clusters is given as\cite{Wohlfarth, Bitoh}
\begin{equation}
M=n\langle\mu\rangle L(\langle\mu\rangle H/k_BT)\label{eq:Wohlfarth}
\end{equation}
where $n$ is the number of clusters per unit volume, $\langle\mu\rangle$ is the average magnetic moment of the clusters, $k_B$ is the Boltzman constant, and $L(x)$ is the Langevin function. Above the blocking temperature ($T_B$), the expansion of Langevin function in powers of $H$ gives
\begin{equation}
\chi_1=n\langle\mu\rangle^2/3k_BT=P_1/T\label{eq:Wohlfarth1}
\end{equation}
and
\begin{equation}
\chi_3=(n\langle\mu\rangle/45)(\langle\mu\rangle/k_BT)^3=P_3/T\label{eq:Wohlfarth2}
\end{equation}
The equation \ref{eq:Wohlfarth1} and \ref{eq:Wohlfarth2} show that for superparamagnetic clusters, above $T_B$, $\chi_1$ and $\chi_3$ varies as a linear function of $T^{-1}$ and $T^{-3}$ respectively. The figure \ref{fig: superpara} show the $\chi^{'}_1$ versus $T^{-1}$ which is linear above $T_B$. Similarly, the inset of figure \ref{fig: superpara} display the linear variation of $\chi^{'}_3$ with $T^{-3}$ above $T_B$. The ratio of the fitting  parameter $P_3$ and $P_1$ is used to estimate $\langle\mu\rangle$, which comes around $1.83\times10^5~\mu_B$ where $\mu_B$ is the effective Bohr magneton. Such a large value of $\langle\mu\rangle$ is generally observed  in superparamagnet clusters. This is because the cluster consists of a large number of atomic spins each having the magnetic moment of few $\mu_B$ (while normal paramagnet only have the atomic spins). Since $\langle\mu\rangle$=$M_SV$ where $M_S$ is the saturation magnetization and $V$ is the volume of cluster, assuming the clusters to be spherical, the average size of the clusters comes around 15~nm. The size of the magnetic clusters is much smaller than the crystallite size ($\approx 85$~nm calculated from the X-ray diffraction) which indicates that each crystallite contains a number ferromagnetic-clusters. Combining this with the results of low temperature isothermal magnetization of section \ref{subsubsection:isothermal magnetization}, we infer that each crystallite of the system consists of percolating ferromagnetic-clusters coexisting along with the non-ferromagnetic hole poor phases.

\begin{figure} [!t]
\begin{centering}
\includegraphics[width=0.8\columnwidth]{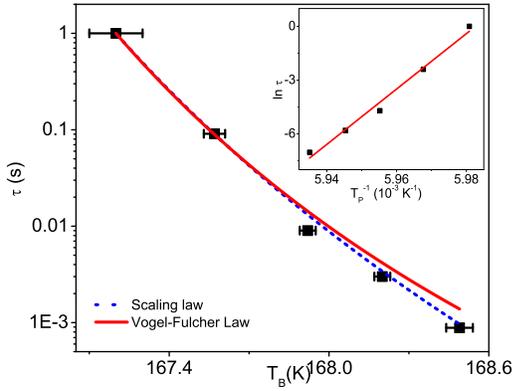}
\par\end{centering}
\caption{Variation of relaxation time ($\tau$) with blocking temperature (T$_B$). The solid line represents the fitting of Vogel-Fulcher law while the dashed line represents the fitting of scaling law. The inset show the ln$\tau$ versus T$_B^{-1}$ and the solid line is the fitting of N\'eel-Arrhennius law.} \label{fig:scaling}
\end{figure}

\subsubsection{Inter-cluster interaction}
The ferromagnetic-clusters coexisting with the non-ferromagnetic phases may interact with each other directly through dipole-dipole interaction or via non-ferromagnetic matrix through exchange interactions.\cite{Bedanta} The degree of inter-cluster interaction and their effect on
fluctuation dynamics is studied by fitting the frequency dependence of $T_B$ with N\'eel-Arrhennius, Vogel-Fulcher, and scaling law.\cite{Binder} The results of these fittings are shown in figure \ref{fig:scaling} and its inset. For an ensemble of non-interacting superparamagnets, the relaxation time $\tau$ follows the N\'eel-Arrhennius  law\cite{Binder}
\begin{equation}
\tau=\tau_0 \text{exp}\left(\frac{E_a}{k_BT}\right)\label{eq:Neel}
\end{equation}
where $E_a$ is the average anisotropy energy barrier, $\tau_0$ is the times constant corresponding to characteristic attempt frequency, and $k_B$ is the Boltzman constant. The experimentally observed $\tau_0$ values for non-interacting super-paramagnets are in the range of $10^{-8}-10^{-13}$~s.\cite{Dormann} The inset of figure \ref{fig:scaling} show the fitting of equation \ref{eq:Neel} to $T_B^{-1}$ versus ln$\tau$ data which gives $\tau_0\approx10^{-402}$~s and $E_a$=13.31~eV. The fitting of N\'eel-Arrhennius law yields un-physical values which rule out the possibility of non-interacting dynamics and hint the presence of cooperative dynamics due to inter-cluster interaction. The dynamics of the interacting superparamagnets is described by Vogel-Fulcher law\cite{Binder}
\begin{equation}
\tau=\tau_0 \text{exp}\left(\frac{E_a}{k_B(T-T_0)}\right)\label{eq:VG}
\end{equation}
where the temperature $T_0$ which has a value between zero and $T_B$ is often related to the strength of inter-cluster interaction. The fitting of equation \ref{eq:VG} to the data of figure \ref{fig:scaling} gives $\tau_0\sim$10$^{-9}$~s, $E_a/k_B$=63(9)~K and $T_0$=164.3~K. The $\tau_0$ value obtained from the Vogel-Fulcher fitting is orders of magnitude larger than the spin-flip time of atomic magnetic moments ($\sim10^{-13}$~s). This strongly supports that the fluctuating entities are spin-clusters with a significant inter-cluster interaction among them.
Strong inter-cluster interactions can give rise to spin-glass like cooperative freezing, and in this case, the frequency dependence of peak in $\chi^{'}_1$ is expected to follow the power law divergence of the standard critical slowing down given by dynamic scaling theory\cite{Maydosh, Binder, Hohenberg}
 \begin{equation}
\tau=\tau_0(T/T_g-1)^{-zv}\label{eq:dscaling}
\end{equation}
where $\tau$ is the dynamical fluctuation time scale corresponding to measurement frequency at the peak temperature of $\chi^{'}_1$, $\tau_0$ is the spin flipping time of the relaxing entities, $T_g$ is the  cluster-glass (or spin-glass) transition temperature in the limit of zero frequency, $z$ is the dynamic scaling exponent, and $v$ is the critical exponent. In the vicinity of cluster-glass transition, the spin cluster correlation length $\xi$ diverges as $\xi\propto(T/T_g-1)^{-v}$ and the dynamic scaling hypothesis relates $\tau$ to $\xi$ as $\tau\sim\xi^{z}$. The fitting of scaling law to the frequency dependence of T$_B$ (see figure \ref{fig:scaling}) gives $\tau_0\sim10^{-25}$~s, $T_g$=165.4~K, and $zv$=12.6. The value of exponent $zv$ is higher than that observed in case of spin-glasses (2-10) and $\tau_0$ is orders of magnitude smaller than the values reported for cluster-glass ($10^{-9}$-$10^{-6}$~s) and spin-glass ($10^{-11}$-$10^{-13}$~s). The value of $\tau_0$ is even smaller than the spin-flip time of a single atom ($\sim10^{-13}$~s), which is un-physical, and this indicates that the spin dynamics in the system does not exhibit the critical slowing down on approaching $T_g$ as expected from the dynamic scaling. Thus, it can be inferred that the inter-cluster interactions present in the system are significant, but not strong enough to cause a spin-glass like transition.

\begin{figure} [!t]
\begin{centering}
\includegraphics[width=0.7\columnwidth]{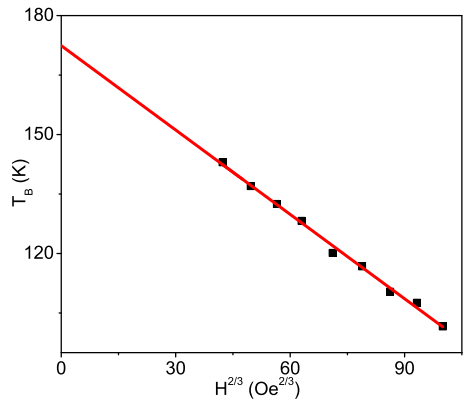}
\par\end{centering}
\caption{Field dependence of the peak in ZFC magnetization. The straight line show the fitting of AT line to data.} \label{fig:ATline}
\end{figure}

\subsection{Further discussions}
The presence of significant inter-cluster interaction can also be reaffirmed from the field dependence of the peak temperature (T$_p$) in ZFC magnetization curves. For Ising spin-glass, mean field theory of spin-glass predicts a critical de Almeida-Thouless (AT) line in $H$-$T$ space which marks the spin-glass phase transition.\cite{Almeida} Above AT line the large field destroys the frozen spin state. The spin-glass transition temperature corresponds to the peak in ZFC magnetization (T$_p$) and the AT line predicts that T$_p\propto H^{2/3}$.  The AT line like field dependence of $T_p$ is not unique to spin or cluster -glass transition but it has been also observed in some of interacting super-paramagnets which otherwise undergo a progressive thermal blocking.\cite{Pramanik, Wenger} In figure \ref{fig:ATline} we have plotted the field dependence of T$_p$ which fits well with the AT line giving zero field spin-glass transition temperature ($T_g$) of 172(1)~K. Since our ac susceptibility measurements have ruled out the possibility of a spin or cluster -glass like freezing, the existence of AT line in $H$-$T$ space clearly indicates the presence of a significant inter-cluster interaction in the system. The inter-cluster interactions can originate from different type of magnetic interactions and the strength of these interactions generally depends on the packing density of ferromagnetic-clusters.
These magnetic interaction includes the long range dipole-dipole interaction among the ferromagnetic-clusters along with the possibilities of exchange, tunneling exchange and superexchange interactions.\cite{Bedanta}

The absence of spin-glass like phase in La$_{0.5}$Ba$_{0.5}$CoO$_3$ is in contrast with La$_{0.5}$Sr$_{0.5}$CoO$_3$ where spin or cluster -glass like phase coexists along with the percolating ferromagnetic-clusters.\cite{Samal1} The doping at A site of LaCoO$_3$ with 50\% Ba or Sr gives same hole concentration, and therefore, the observed discrepancy in magnetically ordered state of La$_{0.5}$Ba$_{0.5}$CoO$_3$ and La$_{0.5}$Sr$_{0.5}$CoO$_3$ can be only due to difference in tolerance factor and local lattice distortions caused by the difference in ionic radii of Ba and Sr. A comparative study  using the microscopic probe e.g neutron scattering is required to understand how this difference
gives rise to non-spin-glass and spin-glass like behavior in these two systems.

\section{Conclusions}
In conclusion, we have performed a comprehensive set of dc magnetization, linear and non-linear ac susceptibility measurements to understand the magnetic state of the hole doped disordered cobaltite La$_{0.5}$Ba$_{0.5}$CoO$_3$. The results of isothermal magnetization suggest that the magnetically ordered state of the system consists of percolating ferromagnetic-clusters along with the coexisting non-ferromagnetic phases. The Kouvel-Fisher analysis of the ac susceptibility gives $T_C$=202(3)~K and $\gamma$=1.29(1) indicating the possibility of 3D Hisenberg like magnetic ordering in the ferromagnetic-clusters. Above $T_C$, there exist a temperature range ($T_C$ <T <$T^*$) where inverse dc susceptibility exhibits a field independent upward deviation from Curie-Weiss law which is different from the field dependent downward deviation seen in Griffiths phase. In this region, the spontaneous magnetization is zero, but the real part of second harmonic of ac susceptibility and the imaginary part of linear ac susceptibility are non-zero which indicate the presence magnetic correlations that do not originate from the variation in local $T_C$ due to composition fluctuation. This suggests that possibly these magnetic correlation comes from the preformation of short range magnetic-clusters.

Below $T_C$ the system exhibits thermomagnetic irreversibility and frequency dependence in the peak of ac susceptibility which suggest the presence of spin-glass, cluster-glass, or superparamagnetic phases. The absence of field divergence in the peak of third harmonic of ac susceptibility and absence of ZFC memory rule out the existence of spin or cluster -glass like phases and  suggest that the observed spin-dynamics is possibly due to super-paramagnetic like thermal blocking of ferromagnetic-clusters. This is in sharp contrast to La$_{0.5}$Sr$_{0.5}$CoO$_3$ where the spin or cluster -glass like phase coexist with the ferromagnetic-clusters. The super-paramagnetic behavior of ferromagnetic-clusters in La$_{0.5}$Ba$_{0.5}$CoO$_3$ is further confirmed by Wohlfarth's model of super-paramagnetism. The analysis of frequency dependence in the peak of ac susceptibility by N\'eel-Arrhennius, Vogel-Fulcher, and scaling law suggest the existence of significant inter-cluster interaction among the ferromagnetic-clusters which is further confirmed by the existence of AT line in the H-T space.

\section{Acknowledgement}
We are thankful to P. Chaddah for useful discussions. We also acknowledge M. Gupta for XRD measurements.


\begin{thebibliography}{30}
\bibitem{Dagotto} E. Dagotto, Science \textbf{309}, 257 (2005).

\bibitem{Rini} M. Rini, R. Tobey, N. Dean, J. Itatani, Y. Tomioka, Y. Tokura, R. W. Schoenlein, and A. Cavalleri, Nature \textbf{449}, 72 (2007).

\bibitem{Louca1} D. Louca, J. L. Sarrao, J. D. Thompson, H. Roder, and G. H. Kwei, Phys. Rev. B \textbf{60}, 10378 (1999).

\bibitem{Podlesnyak} A. Podlesnyak, S. Streule, J. Mesot, M. Medarde, E. Pomjakushina, K. Conder, A. Tanaka, M. W. Haverkort, and D. I. Khomskii, Phys. Rev. Lett. \textbf{97},  247208 (2006).

\bibitem{Raccah} P. M. Raccah and J. B. Goodenough, Phys. Rev. B \textbf{155}, 932 (1967).

\bibitem{Kriener} M. Kriener, M. Braden, H. Kierspel, D. Senff, O. Zabara, C. Zobel, and T. Lorenz, Phys. Rev. B \textbf{79}, 224104 (2009).

\bibitem{Zobel} C. Zobel, M. Kriener, D. Bruns, J. Baier, M. Gr\"uninger, T. Lorenz, P. Reutler, and A. Revcolevschi, Phys. Rev. B \textbf{66}, 020402(R) (2002).

\bibitem{Noguchi} S. Noguchi, S. Kawamata, K. Okuda, H. Nojiri, and M. Motokawa, Phys. Rev. B \textbf{66}, 094404 (2002).

\bibitem{Haverkort} M. W. Haverkort, Z. Hu, J. C. Cezar, T. Burnus, H. Hartmann, M. Reuther, C. Zobel, T. Lorenz, A. Tanaka, N. B. Brookes, H. H. Hsieh, H.-J. Lin, C. T. Chen, and L. H. Tjeng, Phys. Rev. Lett. \textbf{97}, 176405 (2006).

\bibitem{Radaelli} P. G. Radaelli and S.-W. Cheong, Phys. Rev. B \textbf{66}, 094408 (2002).

\bibitem{Korotin} M. A. Korotin, S. Yu. Ezhov, I. V. Solovyev, V. I. Anisimov, D. I. Khomskii, and G. A. Sawatzky, Phys. Rev. B \textbf{54}, 5309 (1996).

\bibitem{Louca} D. Louca and J. L. Sarrao, Phys. Rev. Lett. \textbf{91}, 155501 (2003).

\bibitem{Phelan} D. Phelan, J. Yu, and D. Louca, Phys. Rev. B \textbf{78}, 094108 (2008).

\bibitem{Podlesnyak1} A. Podlesnyak, M. Russina, A. Furrer, A. Alfonsov, E. Vavilova, V. Kataev, B. B\"uchner, Th. Strassle, E. Pomjakushina, K. Conder, and D. I. Khomskii, Phys. Rev. Lett. \textbf{101}, 247603 (2008).

\bibitem{Podlesnyak2} A. Podlesnyak, G. Ehlers, M. Frontzek, A. S. Sefat, A. Furrer, Th. Strassle, E. Pomjakushina, K. Conder, F. Demmel, and D. I. Khomskii, Phys. Rev. B \textbf{83}, 134430 (2011).

\bibitem{Kriener1} M. Kriener, C. Zobel, A. Reichl, J. Baier, M. Cwik, K. Berggold, H. Kierspel, O. Zabara, A. Freimuth, and T. Lorenz, Phys. Rev. B \textbf{69}, 094417 (2004).

\bibitem{Mandal} P. Mandal, P. Choudhury, S. K. Biswas, and B. Ghosh, Phys. Rev. B \textbf{70}, 104407 (2004).

\bibitem{Wu} J. Wu and C. Leighton, Phys. Rev. B \textbf{67}, 174408 (2003).

\bibitem{Mukherjee} S. Mukherjee, R. Ranganathan, P. S. Anilkumar, and P. A. Joy, Phys. Rev. B \textbf{54}, 9267 (1996).

\bibitem{Samal} D. Samal and P. S. Anil Kumar, J. Phys.: Condens. Matter \textbf{23}, 016001 (2011).

\bibitem{Samal1} D. Samal and P. S. Anil Kumar, J. Appl. Phys. \textbf{111}, 043902 (2012).

\bibitem{Phelan1} D. Phelan, D. Louca, K. Kamazawa, M. F. Hundley, and K. Yamada, Phys. Rev. B \textbf{76}, 104111 (2007).

\bibitem{Yu} J. Yu, D. Louca, D. Phelan, K. Tomiyasu, K. Horigane, and K. Yamada, Phys. Rev. B \textbf{80}, 052402 (2009).

\bibitem{Pati} R. K. Pati, J. C. Ray, and P. Pramanik, J. Am. Ceram. Soc. \textbf{84}, 2849 (2001).

\bibitem{Bajpai} A. Bajpai and A. Banerjee, Rev. Sci. Instrum. \textbf{68}, 4075 (1997).

\bibitem{Carvajal} J. Rodr\'iguez-Carvajal, Phys. B (Amsterdam) \textbf{192}, 55 (1993); Program {\small FULLPROF}, LLB-JRC, Laboratoire L\'eon Brillouin, CEA-Saclay, France, 1996.

\bibitem{Troyanchuk}  I. O. Troyanchuk, N. V. Kasper, D. D. Khalyavin, A. N. Chobot, G. M. Chobot, and H. Szymczak, J. Phys.: Condens. Matter \textbf{10}, 6381 (1998).

\bibitem{Maydosh} J. A. Mydosh, Spin Glasses: An Experperimental
Introduction (Taylor and Francis London 1993).


\bibitem{Deac} I. G. Deac, J. F. Mitchell, and P. Schiffer, Phys. Rev. B \textbf{63}, 172408
(2001).

\bibitem{Huang} X. H. Huang, J. F. Ding, Z. L. Jiang, Y. W. Yin, Q. X. Yu, and X.
G. Lia, J. Appl. Phys. \textbf{106}, 083904 (2009).

\bibitem{Knobel} M. Knobel, W. C. Nunes, L. M. Socolovsky, E. De Biasi, J. M. Varg
as, and J. C. Denardin, J. Nanosci. Nanotechnol. \textbf{8}, 2836
(2008).

\bibitem{Pramanik} A. K. Pramanik and A. Banerjee, Phys. Rev. B \textbf{82}, 094402 (2010).

\bibitem{Anil} P. S. Anil Kumar, P. A. Joy, and S. K. Date, J. Phys. : Condens. Matter
\textbf{10}, L487 (1998).

\bibitem{Caciuffo} R. Caciuffo, J. Mira, J. Rivas, M. A. Senaris-Rodriguez, P. G. Radaelli, F. Carsughi, D. Fiorani, and J. B. Goodenough, Europhys. Lett. \textbf{45}, 399 (1999).

\bibitem{He} C. He, M. A. Torija, J. Wu, J. W. Lynn, H. Zheng, J. F. Mitchell, and C. Leighton, Phys. Rev. B \textbf{76}, 014401 (2007).

\bibitem{Griffiths} R. B. Griffiths, Phys. Rev. Lett. \textbf{23}, 17 (1969).



\bibitem{Magen} C. Magen, P. A. Algarabel, L. Morellon, J. P. Ara\'ujo, C. Ritter, M. R. Ibarra, A. M. Pereira, and J. B. Sousa, Phys. Rev. Lett. \textbf{96}, 167201 (2006).

\bibitem{Pramanik1} A. K. Pramanik and A. Banerjee, Phys. Rev. B \textbf{81}, 024431 (2010).

\bibitem{Ravindran} P. Ravindran, H. Fjellvag, A. Kjekshus, P. Blaha, K. Schwarz, and J. Luitz, J. Appl. Phys. \textbf{91}, 291 (2002).

\bibitem{Arrott} A. Arrott, Phys. Rev. \textbf{108}, 1394 (1957).

\bibitem{Goya} G. F. Goya and V. Sagredo, Phys. Rev. B \textbf{64}, 235208 (2001).

\bibitem{Cong} D. Y. Cong, S. Roth, J. Liu, Q. Luo, M. P\"otschke, C. H\"urrich, and L. Schultz, Appl. Phys. Lett. \textbf{96}, 112504 (2010).

\bibitem{Thakur} M. Thakur, M. Patra, S. Majumdar, and S. Giri, J. Appl. Phys. \textbf{105}, 073905 (2009).

\bibitem{Toro} J. A. De Toro, M. A. L\'opez de la Torre, M. A. Arranz, J. M. Riveiro, J. L. Mart\'inez, P. Palade, and G. Filoti, Phys. Rev. B \textbf{64}, 094438 (2001).

\bibitem{Dormann} J. L. Dormann, L. Bessais, and D. Fiorani, J. Phys. C \textbf{21}, 2015 (1988).

\bibitem{Kouvel} J. S. Kouvel and M. E. Fischer, Phys. Rev. \textbf{136}, A1626 (1964).

\bibitem{Menyuk} N. Menyuk, P. M. Raccah, and K. Dwight, Phys. Rev. \textbf{166}, 510 (1968).

\bibitem{Mukherjee} S. Mukherjee, P. Raychaudhuri, and A. K. Nigam, Phys. Rev. B \textbf{61}, 8651 (2000).

\bibitem{Khan} N. Khan, A. Midya, K. Mydeen, P. Mandal, A. Loidl, and D. Prabhakaran, Phys. Rev. B \textbf{82}, 064422 (2010).

\bibitem{Suzuki} M. Suzuki, Prog. Theor. Phys. \textbf{58}, 1151 (1977).

\bibitem{Bajpai1} A. Bajpai and A. Banerjee, Phys. Rev. B \textbf{55}, 12439 (1997).

\bibitem{Wohlfarth} E. P. Wohlfarth, J. Appl. Phys. \textbf{29}, 595 (1958).

\bibitem{Bitoh} T. Bitoh, K. Ohba, M. Takamatsu, T. Shirane, and S. Chikazawa, J. Magn. Magn. Mater. \textbf{154}, 59 (1996).

\bibitem{Bedanta} S. Bedanta and W. Kleemann,  J. Phys. D: Appl. Phys. \textbf{42}, 013001 (2009).

\bibitem{Binder} K. Binder and A. P. Young, Rev. Mod. Phys. \textbf{58}, 801 (1986).

\bibitem{Hohenberg} P. C. Hohenberg and B. I. Halperin, Rev. Mod. Phys. \textbf{49}, 435 (1977).

\bibitem{Dormann} J. L. Dormann, D. Fiorani, and E. Tronc, J. Magn. Magn. Mater. \textbf{202}, 251 (1999).

\bibitem{Djurberg} C. Djurberg, P. Svedlindh, P. Nordblad, M. F. Hansen, F. Bodker, and S. Morup, Phys. Rev. Lett. \textbf{79}, 5154 (1997).

\bibitem{Almeida} J. R. L. de Almeida and D. J. Thouless, J. Phys. A \textbf{11}, 983 (1978).

\bibitem{Wenger} L. E. Wenger and J. A. Mydosh, Phys. Rev. B \textbf{29}, 4156 (1984).

\end{thebibliography}
\end{document}